\documentclass[lettersize,journal]{IEEEtran}
\usepackage{amsmath,amsfonts}
\usepackage{algorithmic}
\usepackage{array}
\usepackage[caption=false,font=normalsize,labelfont=sf,textfont=sf]{subfig}
\usepackage{textcomp}
\usepackage{stfloats}
\usepackage{makecell}  
\usepackage{url}
\usepackage{verbatim}
\usepackage{graphicx}
\usepackage{multirow}  
\usepackage{colortbl}  
\usepackage{amsmath}   
\usepackage{array}     
\usepackage{amssymb}   
\usepackage{pifont}    
\usepackage{cite}
\def\BibTeX{{\rm B\kern-.05em{\sc i\kern-.025em b}\kern-.08em
    T\kern-.1667em\lower.7ex\hbox{E}\kern-.125emX}}
\usepackage{balance}
\begin{document}
\title{Iterative Collaboration Network Guided By\\ Reconstruction Prior for Medical Image Super-Resolution}
\author{Xiaoyan Kui, Zexin Ji, Beiji Zou, Yang Li, Yulan Dai, Liming Chen, \textit{Senior Member
, IEEE}, Pierre Vera, and Su Ruan, \textit{Senior Member, IEEE}
\thanks{The work was supported by the National Key R$\&$D Program of China (No.2018AAA0102100); National Natural Science Foundation of China (Nos.U22A2034, 62177047); High Caliber Foreign Experts Introduction Plan funded by MOST; Key Research and Development Program of Hunan Province (No.2022SK2054); Major Program from Xiangjiang Laboratory under Grant 23XJ02005; Key Research and Development Programs of Department of Science and Technology of Hunan Province(No.2024JK2135); Scientific Research Fund of Hunan Provincial Education Department (No. 24A0018); Central South University Research Programme of Advanced Interdisciplinary Studies (No.2023QYJC020); Natural Science Foundation of Hunan Province (No.2024JJ6338); Fundamental Research Funds for the Central Universities of Central South University (No.2024ZZTS0486);  China Scholarship Council (No.202306370195).(Corresponding author: Zexin Ji.)

Zexin Ji is with the School of Computer Science and Engineering, Central South University, Changsha, 410083, China, and also with the Department of Nuclear Medicine, Henri Becquerel Cancer Center, Rouen, France.
(e-mail:zexin.ji@csu.edu.cn).

Xiaoyan Kui, Beiji Zou and Yulan Dai are with the School of Computer Science and Engineering, Central South University, Changsha, 410083, China.

Yang Li is with the School of Informatics, Hunan University of Chinese Medicine, Changsha, 410208, China.

Liming Chen is with the Department of Mathematics and Computer Science, Ecole Centrale de Lyon, Ecully 69130, France.

Pierre Vera is with the Department of Nuclear Medicine, Henri Becquerel Cancer Center, Rouen, France.

Su Ruan is with the University of Rouen-Normandy, AIMS - QuantIF UR 4108, F-76000, Rouen, France.

}}

\markboth{Journal of \LaTeX\ Class Files,~Vol.~18, No.~9, June~2024}%
{How to Use the IEEEtran \LaTeX \ Templates}

\maketitle

\begin{abstract}

High-resolution medical images can provide more detailed information for better diagnosis. Conventional medical image super-resolution relies on a single task which first performs the extraction of the features and then upscaling based on the features.
The features extracted may not be complete for super-resolution. Recent multi-task learning,
including reconstruction and super-resolution, is a good solution to obtain additional relevant information.
The interaction between the two tasks is often insufficient, which still leads to incomplete and less relevant deep features.
To address above limitations, we propose an iterative collaboration
network (ICONet) to improve communications between
tasks by progressively incorporating reconstruction prior
to the super-resolution learning procedure in an iterative
collaboration way. It consists of a reconstruction branch, a super-resolution branch, and a SR-Rec fusion module. 
The reconstruction branch generates the artifact-free image as prior, which is followed by a super-resolution branch for prior knowledge-guided super-resolution. Unlike the widely-used convolutional neural networks for extracting local features and Transformers with quadratic computational complexity for modeling long-range dependencies, we develop a new residual spatial-channel feature learning (RSCFL) module of two branches to efficiently establish feature relationships in spatial and channel dimensions. Moreover, the designed SR-Rec fusion module fuses the reconstruction prior and super-resolution features with each other in an adaptive manner. Our
ICONet is built with multi-stage models to iteratively
upscale the low-resolution images using steps of ${2 \times}$ and
simultaneously interact between two branches in multi-stage supervisions. Quantitative and qualitative experimental results on the benchmarking dataset show that our ICONet outperforms most state-of-the-art approaches.

\end{abstract}


\section{Introduction}
\IEEEPARstart{M}{edical} imaging is an indispensable tool in medical research and clinical applications. High-resolution medical images, which provide more detailed information, require longer scanning times, as well as expensive equipment. In clinical settings, there is a significant demand for high-resolution medical images despite constraints in scanning time and hardware capabilities. Consequently, enhancing the resolution of medical images is crucial. The image super-resolution (SR) has gained significant attention in medical image processing due to its cost-effectiveness. Specifically, medical image super-resolution aims to generate high-resolution~(HR) images from their low-resolution~(LR) counterparts, offering a practical solution to improve image quality without the need for more advanced hardware and longer scanning time.

\begin{figure}[t] 
	\centering
	\includegraphics[width=1\linewidth]{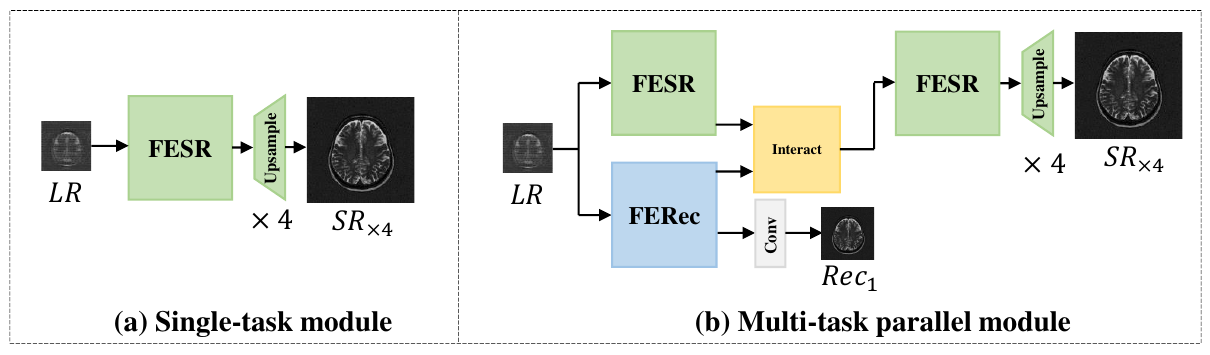}
	\caption{The $4 \times$ super-resolution module examples of (a) the single task module and (b) the multi-task parallel module. \textbf{FESR}: feature extraction for super-resolution task. \textbf{FERec}: feature extraction for reconstruction task. \textbf{Interact}: Interaction module between two tasks.	
	}
	\label{fig:motivation}  
\end{figure}

The emergence of deep learning techniques has significantly enhanced the field of image super-resolution~\cite{DBLP:journals/tip/ChenZZ24,li2023hyperspectral,DBLP:journals/tcyb/ZhangSDZ21}, primarily due to the advanced generative capabilities of convolutional neural networks (CNNs). Currently, CNN-based medical image super-resolution methods~\cite{8,9,10,11,12,13,DBLP:journals/nn/SongCYD20,DBLP:journals/tci/WanWWGSW24} have demonstrated improvements over traditional super-resolution approaches~\cite{14,15}. As an example, the Super-resolution Convolutional Neural Network (SRCNN)~\cite{8} pioneered the use of CNNs for super-resolution challenges. Kim \emph{et al.}~\cite{9} extended the CNN's depth to 20 layers, enhancing its ability to represent robust features. Qiu \emph{et al.}~\cite{10} combined SRCNN with a sub-pixel convolutional layer, which significantly enhanced the quality of MR images. The TransMRSR network~\cite{DBLP:journals/vc/HuangLTHWCS23} employs convolutional blocks for local information and Transformer blocks for global details to realize the brain image super-resolution. These methods can be regarded as the single-task module (Fig.\ref{fig:motivation} (a)), where a LR image is input, processed through a feature extraction network, and then output as a super-resolved (SR) image. However, these methods only exploit the knowledge of single task, ignoring the auxiliary information of other image generation tasks. 
To solve this issue, multi-task learning methods have been introduced that utilize auxiliary tasks to enhance medical image super-resolution. For example, a task transformer network~(T$^2$Net)~\cite{12} was designed for joint MRI reconstruction and super-resolution (Fig.\ref{fig:motivation} (b)). Specifically, the reconstruction task outputs an artifact-free image with the same scale as the original. Super-resolution task outputs high-resolution images, e.g. 4$\times$ upsampling. T$^2$Net can share the feature representation of multiple tasks and utilize the reconstruction information to super-resolution task in a parallel manner. 
However, the current joint reconstruction and super-resolution methods also have certain limitations. 1) It struggles to fully exploit the auxiliary information of reconstruction task because the reconstruction process is conducted on low-resolution input images with the same scale of output. Consequently, single-scale reconstruction guidance may be limited for multi-scale upsampling in super-resolution task.
2) The widely used feature extraction modules, such as ResNet~\cite{DBLP:conf/cvpr/HeZRS16} and Transformer~\cite{DBLP:conf/iclr/DosovitskiyB0WZ21}, fail to efficiently mine effective features beneficial for image super-resolution. Specifically, ResNet can only model local features, and Transformer, which can model long-range dependencies, has a quadratic computational complexity.
3) Most methods primarily focus on optimizing the super-resolution network on a single scale. However, it can not provide the optimal supervision signals for efficient training since the details across different scales are not completely captured and utilized. 
Therefore, it is essential to adequately mine the auxiliary information of reconstruction to further improve super-resolution performance.

To address the aforementioned problems, we propose an iterative collaboration method for medical image super-resolution in line with (Fig.\ref{fig:Framework}). Motivated by the fact that the super-resolution task can obtain high-resolution images under the guidance of  reconstruction prior and the reconstruction module also benefits from high-resolution inputs. To achieve this, we design an iterative collaboration network~(ICONet), which is a multi-stage iterative architecture instead of a very deep model for super-resolution. The previous output of the reconstruction branch is fed into the super-resolution branch in an iterative collaboration way, so that two branches collaborate with one another for improved performance. 
Recently, State Space Models (SSMs)\cite{DBLP:conf/nips/GuJGSDRR21} have emerged in deep learning as a promising architecture for sequence modeling, which provides an alternative to CNNs and transformers for capturing long-range dependencies. Unlike Transformers that require large parameter scales, SSMs offer the advantage of linear scaling with sequence length.
Building on structured state space sequence (S4) model~\cite{DBLP:conf/iclr/GuGR22}, a new SSM architecture named Mamba~\cite{DBLP:journals/corr/abs-2312-00752} was recently introduced. The Mamba developed a selective scan state space sequential model, which is a data-related SSM with selective mechanisms and efficient hardware design. The advancements of Mamba in sequence modeling have significantly sparked interest in the computer vision community~\cite{xing2024segmamba,DBLP:journals/corr/abs-2402-02491,zhu2024vision,DBLP:conf/cvpr/WangZWYLOH23,DBLP:journals/corr/abs-2401-10166}. Inspired by the success of Mamba, we design the new residual spatial-channel feature learning (RSCFL) module as the feature extraction form to efficiently mine long-range dependency features on the spatial and channel dimension. 
For spatial feature learning, we use vision Mamba block (VMB) to efficiently model long-distance dependencies of medical image content. For channel feature learning, we design the channel ranking block (CRB) to focus more on foreground features with richer information in medical images. The two sequential blocks constitute a residual learning block module. Furthermore, a SR-Rec fusion module as the interaction is set up to effectively integrate reconstruction prior with the super-resolution task. Moreover, the multi-scale supervised loss function can further extract different detailed information from the image. 
The quantitative and qualitative results show that our proposed ICONet outperforms other state-of-the-art super-resolution approaches.

The contributions of our ICONet are:
\begin{itemize}
	\item[$\bullet$] Developing a new iterative collaboration network~(ICONet) that iteratively reasons out high-resolution medical images with the guidance from reconstruction prior. 
	\item[$\bullet$] Designing a residual spatial-channel feature learning (RSCFL) module that can efficiently extract relevant features on the spatial and channel dimensions. An effective SR-Rec fusion module that adaptively fuses the reconstruction prior to the super-resolution task at every stage.
	\item[$\bullet$] Breaking down the challenging high-upsampling super-resolution task into sub-tasks to progressively restore the low-resolution image with multi-stage supervision.
	\item[$\bullet$] Achieving good performance on the MRI and PET benchmarks to confirm the effectiveness of proposed ICONet. Furthermore, we also provide detailed quantitative, qualitative, and ablation analysis.
\end{itemize}

\section{Related work}\label{Related_Work}
This section provides a brief overview of current medical image super-resolution methods, particularly those using prior knowledge-based super-resolution methods.
\subsection{Medical Image Super-Resolution}
Existing medical image SR approaches can be classified into two categories: traditional and deep learning-based methods. Traditional algorithms contain interpolation-based~\cite{1}, reconstruction-based~\cite{2,14}, and example-based methods~\cite{17}. Interpolation-based SR approaches, such as bicubic and b-spine, are straightforward and speedy but have problems with accuracy. 
Reconstruction-based SR techniques~\cite{2} use the prior knowledge of the input image to limit the range of potential solution space. Example-based methods~\cite{4} use machine learning algorithms to obtain statistical relationships between LR and HR images. Specifically, Yang \emph{et al.}~\cite{5} applied sparse coding methods to solve SR problems. The performance of these methods deteriorates quickly as the upsampling factor rises. In recent years, the deep-learning technique has been adopted to image super-resolution task~\cite{8,9,10,11,12,13,DBLP:journals/nn/SongCYD20,DBLP:journals/tci/LiuLZWL24,DBLP:journals/tci/ZhangCLLRZ25}, which learns discriminative features to model the complex nonlinear relationship between LR and HR image. For example, Dong \emph{et al.}~\cite{8} firstly designed a three-layer convolutional neural network~(CNN) to learn the LR to HR mapping. For more accurate feature representation, Lim \emph{et al.}~\cite{13} used residual block to design an enhanced deep SR network~(EDSR). Several methods have now been applied to the SR task of medical images. Zhang \emph{et al.}~\cite{19} designed an increased convolution layer for MR images super-resolution. Particularly, Qiu \emph{et al.}~\cite{10} adopted SRCNN and a pixelshuffle layer to obtain high-quality medical image. The generative adversarial network~(GAN)~\cite{22} was also employed to super-resolve the low-resolution MR image. Song \emph{et al.}~\cite{DBLP:journals/nn/SongCYD20} utilized the GAN for positron emission tomography (PET) super-resolution. However, CNNs inherently struggle with modeling global contextual information, which is critical in capturing the complex anatomical structures present in medical images.
To address these limitations, Transformer-based approaches have recently gained traction in medical image super-resolution. Regarding the transformer, Huang~\cite{DBLP:conf/aaai/HuangCYWL24} introduced a multi-level feature transfer network that combines a pyramid-based reconstruction network and multi-scale feature extraction to enhance detail recovery in low-resolution MRI images. It further employs MRI-Transformer modules and a contrastive learning constraint to improve feature alignment and texture restoration using reference images. 
However, Transformer typically exhibit quadratic complexity with respect to input sequence length, which can become a limiting factor in high-resolution medical imaging tasks.
Furthermore, the aforementioned medical image SR methods exclusively take into account information relevant to their single task, omitting any auxiliary data from other image generation tasks.

\subsection{Prior knowledge-based super-resolution}
Numerous image priors are introduced to the super-resolution model to improve the performance, named prior knowledge-based super-resolution methods. In multi-contrast MRI SR~\cite{24,DBLP:conf/bibm/JiKLZLDZ23}, different contrast MR images can be regarded as a prior assisted super-resolution. Zheng \emph{et al.}~\cite{24} utilized comparable local regression weights from the multi-contrast MR images to improve the MR image resolution. Zou \emph{et al.}~\cite{DBLP:journals/bspc/ZouJZDZK23} utilized high-quality proton density weighted images~(PDWI) as prior to guide the super-resolution of fat-suppressed proton density weighted images~(FS-PDWI).
In face super-resolution~\cite{26}, facial prior~(facial landmark and parsing map) can be used to improve the face image quality. For example, Ma \emph{et al.}~\cite{26} generated high-resolution face images by using the prior knowledge of landmarks, which in turn allows for more precise landmark localization. Chen \emph{et al.}~\cite{27} restored low-resolution inputs through facial parsing map transformation, which also proposed the new loss function to better synthesize details. Overall, the multi-contrast SR methods depend on different contrast images of the same object. Face prior SR methods depend on accurate prior information specific to the face.
While these prior knowledge-based methods promote the development of super-resolution, these methods heavily rely on extrinsic references. To solve these problems, Feng \emph{et al.}~\cite{12} proposed the T$^2$Net architecture, which jointly realizes MRI super-resolution and reconstruction. Specifically, T$^2$Net uses the reconstructed knowledge, rather than specific prior knowledge, to make access to prior simpler. However, this method lacks the sufficient exploitation of the complementary relationship between two tasks.

\subsection{Preliminaries of Mamba}
Mamba has recently gained attention as an effective method for modeling long-range dependencies. Unlike Transformers, which have computational complexity that scales quadratically with sequence length, Mamba operates with linear complexity. This model is used to define state representations and predict their future states from given inputs. It transforms a one-dimensional function or sequence $x(t) \in \mathbb{R}$ to produce an output $y(t) \in \mathbb{R}$, utilizing a hidden state $h(t) \in \mathbb{R}^{\mathbb{N}}$. The transformation typically occurs through linear ordinary differential equations (ODEs).

\begin{equation}
    h^{\prime}(t)=\mathbf{A} h(t)+\mathbf{B} x(t), y(t)=\mathbf{C} h(t),
    \label{eq:state_space}
\end{equation}
where $\mathbf{A} \in \mathbb{R}^{\mathrm{N} \times \mathrm{N}}$ is the state matrix. $\mathbf{B} \in \mathbb{R}^{\mathbb{N} \times 1}$ and $ \mathbf{C} \in \mathbb{R}^{1 \times \mathbb{N}}$ are the projection parameters.

The zero-order hold (ZOH) method~\cite{DBLP:journals/tcas/GaliasY08} transforms continuous ordinary differential equations (ODEs) into discrete functions, enhancing their compatibility with deep learning frameworks. This model introduces a timescale parameter $\Delta$, which assists in translating the continuous-time system matrices $\mathbf{A}$ and $\mathbf{B}$ into their discrete counterparts $\overline{\mathbf{A}}$ and $\overline{\mathbf{B}}$. The discretization process unfolds as follows:

\begin{equation}\label{eq2}
\overline{\mathbf{A}}=\exp (\Delta \mathbf{A}), \overline{\mathbf{B}}=(\Delta \mathbf{A})^{-1}(\exp (\Delta \mathbf{A})-\mathbf{I}) \cdot \Delta \mathbf{B} .
\end{equation}

After discretization, Equation~\ref{eq:state_space} is transformed into a format appropriate for processing in discrete time, as follows:
\begin{equation}\label{eq3}
h_t=\overline{\mathbf{A}} h_{t-1}+\overline{\mathbf{B}} x_t, y_t=\mathbf{C} h_t .
\end{equation}

The recent progress of Mamba in sequence modeling has drawn considerable attention within the computer vision field~\cite{DBLP:journals/corr/abs-2401-04722,DBLP:conf/miccai/JiZKVR24,DBLP:journals/prl/JiZKLVR25}. For instance, Ma~\emph{et al.}~\cite{DBLP:journals/corr/abs-2401-04722} adopted an encoder-decoder architecture and introduced U-Mamba for biomedical image segmentation. 
Ji~\emph{et al.}~\cite{DBLP:conf/miccai/JiZKVR24} proposed Deform-Mamba framework integrates deformable and Mamba modules to efficiently capture local-global features for MR image super-resolution. They also
designed an improved 2D-Selective-Scan~\cite{DBLP:journals/prl/JiZKLVR25} to adaptively fuse multi-directional long-range dependencies, and edge-aware constraints for improved texture boundary generation. Distinct from previous efforts, this paper is among the first to investigate the potential of Mamba in the context of medical image reconstruction and super-resolution.

\begin{figure*}[t]
    \centering
    \includegraphics[width=1\linewidth]{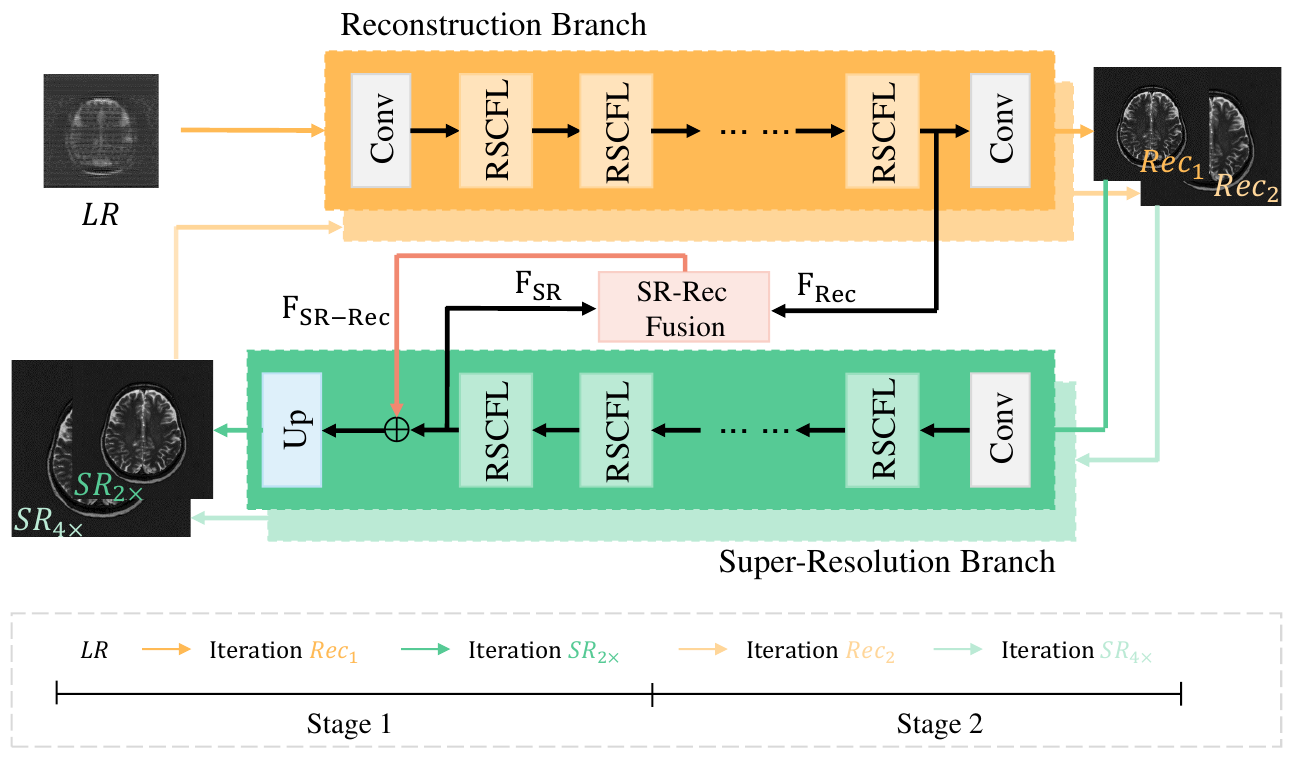}
    \caption{The pipeline of the proposed iterative collaboration network~(ICONet). The main objective of our ICONet are three-fold: 1) the artifacts-free output of the reconstruction branch is exploited as the prior information in the super-resolution branch; 2) the super-resolution branch super-resolves the artifacts-free low-resolution image to high-resolution image with the prior information based on reconstruction; 3) the SR-Rec fusion module is designed to strengthen the guidance of reconstruction prior and effectively fuse the prior to the super-resolution task. Furthermore, we progressively build an iterative collaboration multi-stage network instead of a very deep model, which effectively collaborates with reconstruction and super-resolution branches and enables a higher upsampling factor. RSCFL means the residual spatial-channel feature learning module.}
    \label{fig:Framework}
\end{figure*}

\section{Method}\label{Method}

\subsection{Overview}

We propose an iterative collaborative network (ICONet) for medical image super-resolution. The main network architecture is shown in Fig.~\ref{fig:Framework}. It can be seen that the ICONet contains a reconstruction branch, a super-resolution branch, and a SR-Rec fusion module. 
Firstly, the input LR is processed through reconstruction branch to generate a reconstructed image $Rec_1$ without artifacts, which can reduce the difficulty of image super-resolution. 
Then $Rec_1$ undergoes a super-resolution branch with 2$\times$ upsampling to generate super-resolved image $SR_{2\times}$. 
Specifically, we develop the residual spatial-channel feature learning (RSCFL) module to serve as the basic feature extraction block of the reconstruction and super-resolution branch. The RSCFL module mainly consists of vision Mamba block (VMB), channel ranking block (CRB), and residual learning. Furthermore, we also ensemble the SR-Rec fusion module to effectively enhance the complementary collaboration between reconstruction and super-resolution information. 
Our main proposal is a multi-stage network that iteratively promotes reconstruction and super-resolution, and progressively learns higher upsampling factors for low-resolution inputs. Unlike methods that rely solely on single-scale reconstruction priors, we re-introduce the intermediate $SR_{2\times}$ result into the reconstruction branch to guide the generation of multi-scale reconstructed features, which in turn facilitates the subsequent $SR_{4\times}$ output. This cross-stage interaction enhances contextual representation and helps mitigate the error accumulation typically found in single-stage networks. The two tasks mutually reinforce each other, enabling the network to gradually correct errors and refine textures across stages. By decomposing the upsampling process into more manageable sub-tasks and introducing hierarchical supervision, ICONet simplifies optimization, improves training stability, and achieves superior reconstruction quality. Details about the iterative collaboration network~(ICONet) are discussed in the following.

\begin{figure}
\centering\includegraphics[width=1\linewidth]{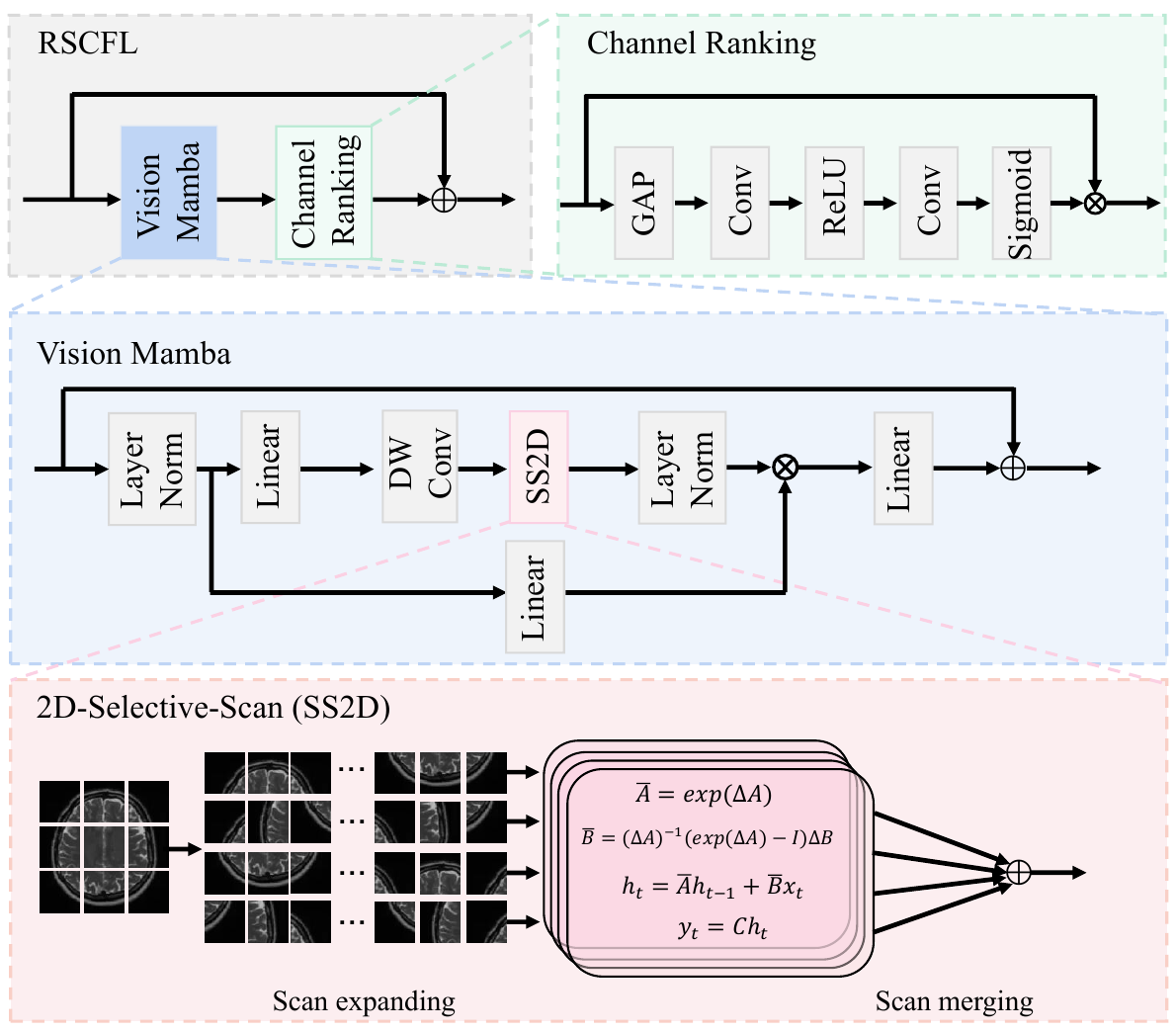}
    \caption{Architecture details of residual spatial-channel feature learning (RSCFL) module.}
    \label{fig:MambaFramework}
\end{figure}

\begin{figure*}[t]
	\centering
	
	\includegraphics[width=1\linewidth]{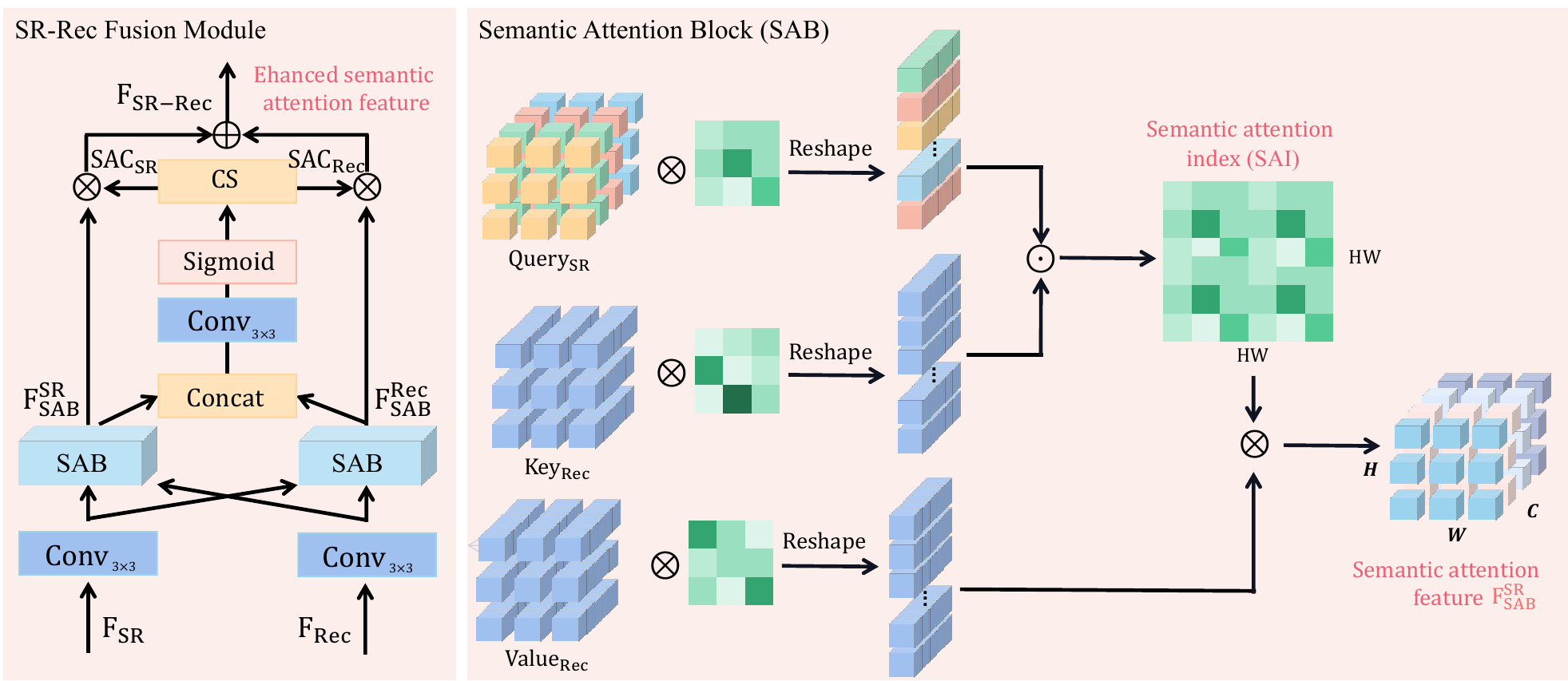}
	\caption{The architecture of the SR-Rec fusion module and semantic attention block (SAB). CS means channel split.
	}
	\label{fig:FusionNet}
\end{figure*}

\subsection{Residual Spatial-Channel Feature Learning Module}
The network structure of residual spatial-channel feature learning (RSCFL) module shown in Fig.~\ref{fig:MambaFramework} consists of a vision Mamba block (VMB) and a channel ranking block (CRB). The two cascaded blocks form a residual learning block module. The VMB is responsible for long-range feature learning at the spatial level, while the CRB is exploring channel level feature ranking.

\textbf{Vision Mamba Block (VMB).}
Similar to the Transformer, Mamba processes its input by dividing the image into a series of patches. As illustrated in Fig.~\ref{fig:MambaFramework}, the vision Mamba module's structure begins with layer normalization of the input patch feature map, which then bifurcates into two paths. In the first path, the input is processed through a linear layer to perform preliminary feature mapping. In the second path, the input passes through a linear layer, followed by a depthwise separable convolution (DW Conv), then a 2D-Selective-Scan (SS2D), and finally another layer normalization. These two parallel paths are merged through element-wise multiplication to effectively emphasize salient features, followed by an additional linear transformation. This combined approach integrates refined context and detailed spatial information into the original input, which significantly enhances the robustness of feature representations. 

The core computational step of this path is the 2D-Selective-Scan (SS2D), designed explicitly to model spatial interactions across image patches. Unlike sequential data in natural language processing, image patches in medical imaging typically lack direct inferential relationships. 
Therefore, we explicitly enhance spatial reasoning among image patches through a scan-expansion process in multiple directional sequences, which aligns with inherent anatomical and structural orientations in medical images. Specifically, the scanning directions include left-to-right, right-to-left, top-to-bottom, and bottom-to-top, to comprehensively model spatial dependencies across horizontal and vertical dimensions.
Each directional scan generates distinct sequential feature representations that are further processed within the hidden state space using specifically designed computational equations. 
These equations facilitate the capture of global dependencies and thus significantly enhance the capability to reason about spatial relationships across distant regions of the image.
Finally, the scan merging process synthesizes these multi-directional sequential features back into a unified representation to restore the original spatial dimension.
This comprehensive directional reasoning approach enables the vision Mamba block (VMB) to model long-range spatial dependencies more effectively and improve performance in medical image super-resolution task.

\textbf{Channel Ranking Block (CRB).}
Unlike natural scene images, medical images contain a large amount of noisy background areas, and physicians are only interested in the information-rich foreground regions. Each feature channel map typically correlates with different semantic attributes. While most channel features capture foreground region-specific features critical for clinical diagnosis, some channels inevitably become dominated by irrelevant background noise. These noisy channels introduce redundant or misleading information, potentially hindering accurate image super-resolution by obscuring subtle but clinically significant features. To mitigate this issue, we aim to selectively minimize the influence of background noise by dynamically emphasizing informative channels and suppressing irrelevant ones through adaptive weighting.

Inspired by the channel attention mechanism introduced in~\cite{DBLP:conf/cvpr/HuSS18}, we propose the channel ranking block (CRB) specifically tailored for medical imaging tasks. As illustrated in Fig.~\ref{fig:MambaFramework}, the CRB dynamically modulates significance of each channel map by explicitly modeling inter-channel dependencies. 
First, global average pooling (GAP) condenses each channel into a scalar to efficiently encode global context. Subsequently, convolutional layers followed by ReLU activation model nonlinear inter-channel dependencies. A Sigmoid function then generates normalized attention weights, adaptively scaling the original channels via element-wise multiplication. Thus, the CRB emphasizes foreground-associated channels while suppressing noisy background-dominated ones, enhancing medical image super-resolution performance.

\subsection{SR-Rec Fusion Module}
The main goal of our SR-Rec fusion module is to complementary fuse the auxiliary information of reconstruction and the super-resolution information to enhance features. Fig. \ref{fig:FusionNet} shows the architecture of our SR-Rec fusion module in detail. The reconstructed features $F_{Rec}$ and super-resolved features $F_{SR}$ are used as inputs to the module. Then, we dynamically fuse two branches by the semantic attention block (SAB) to obtain the SR-Rec semantic attention feature, respectively. To emphasize the semantic relationship between super-resolution and reconstruction, we firstly map the SR input into $Query_{SR}$, the Rec input into $Key_{Rec}$, $Value_{Rec}$ for the super-resolution branch. Then, we input them into the SAB to obtain the semantic attention relevance $R_{SR,Rec}(i,j)$ by calculating the semantic similarity between the $Query_{SR}$ and $Key_{Rec}$.

\begin{equation}
\begin{aligned}
R_{SR,Rec}(i,j) = \frac{Query_{SR}^i}{\Vert Query_{SR}^i \Vert} \odot \frac{Key_{Rec}^j}{\Vert Key_{Rec}^j \Vert},
\end{aligned}
\end{equation}
where $\odot$ is the inner product operation. $i$ and $j$ are the corresponding position in $Query_{SR}$ and $Key_{Rec}$, respectively. The value in  $R_{SR,Rec}(i,j)$ stands for the semantic attention score between the $Query_{SR}^i$ and $Key_{Rec}^j$. The $R_{SR,Rec}(i,j)$ is further used to calculate the semantic attention index (SAI). Specifically, the SAI can obtain the most attention position corresponding to the SR in the $Value_{Rec}$, which can accurately transfer the effective feature in the reconstruction process. It can be calculated by 
\begin{equation}
SAI=\underset{Rec^j}{\operatorname{argmax}} R_{SR,Rec}(i,j),
\end{equation}
To obtain the transferred semantic attention features, we apply SAI to the unfold features of $Value_{Rec}$:
\begin{equation}
F_{SAB}^{SR}=Value_{Rec} \otimes SAI.
\end{equation}

For the features of reconstruction branch, we put Rec input into $Query_{Rec}$, the SR input into $Key_{SR}$, $Value_{SR}$ for the same processing to get $F_{SAB}^{Rec}$. To fully incorporate the information between $F_{SAB}^{SR}$ and $F_{SAB}^{Rec}$, the effective reconstruction feature should be strengthened, which can avoid interference of invalid information. To achieve this,
we concatenate two features and obtain the semantic attention confidence (SAC) through the convolutional layer and sigmoid. Then we split the features along the channel dimension to generate task-specific $SAC_{SR}$ and $SAC_{Rec}$.
Finally, the transferred semantic attention features are element-wisely multiplied by the corresponding SAC to generate the enhanced semantic attention feature. The final output of the module can be interpreted as:
\begin{equation}
F_{SR-Rec}=(F_{SAB}^{SR} \otimes SAC_{SR}) \oplus (F_{SAB}^{Rec} \otimes SAC_{Rec}) .
\end{equation} 
In summary, the SR-Rec fusion module can mutually fuse the artifacts-free feature of reconstruction and the super-resolution feature, which can further boost the feature representation of the super-resolution branch.

\subsection{Multi-Stage Loss Function}
We also introduce a multi-stage loss function to
optimize our ICONet. It is important to provide ground-truth supervision at each stage for progressive super-resolution. Compared with a direct supervision constraint at the final scale, multi-stage supervision can better control the training process and implement higher magnification upsampling. During the training process, the parameters of both reconstruction and super-resolution branches are learned in a mutual-boosted manner.
Specifically, $\mathcal{L}_1$ loss is preferable to $\mathcal{L}_2$ loss in 
terms of boosting performance and convergence~\cite{44}. We utilize $\mathcal{L}_1$ loss function to measure the difference between the network prediction and the ground truth image, and the  multi-stage loss function can be expressed as:

\begin{equation}
\begin{split}
Loss = \alpha \mathcal{L}_{SR} + (1-\alpha) \mathcal{L}_{Rec} = \alpha\sum_{i=1}^{n}{\Vert SR_i-HR_i \Vert}_1 + \\ (1-\alpha)\sum_{i=1}^{n}{\Vert Rec_i-HR_i \Vert}_1.
\end{split}
\end{equation}
where $SR_i$ is the super-resolved MR image in the stage $i$, $Rec_i$ is the reconstructed MR image in the stage $i$, and $HR_i$ is the $i$-th stage ground-truth high-resolution image. $n$ is the stage number. $\alpha$ weights the trade-off between the super-resolution task and reconstruction task, which is selected by weight combination search (Tab.\ref{tab7}).

\section{Experiments}\label{Experiments}
\subsection{Datasets and Evaluation Metrics}
We validated the effectiveness of our ICONet using T2 weighted brain MR images from the IXI dataset\footnote{http://brain-development.org/ixi-dataset/} and FS-PD weighted knee MR images from the fastMRI dataset\footnote{https://fastmri.med.nyu.edu/}. We also verified it using the PET image from the Hecktor dataset \footnote{https://hecktor.grand-challenge.org/}. The downsampling ratio is set to 2 and 4.

\textbf{IXI.}
The IXI dataset mainly contains MR normal brain images of healthy subjects collected from three different hospitals in London. We used 368 subjects for training and 92 subjects for testing in the IXI dataset. The width and height of the brain MRI data are both 256. 

\textbf{fastMRI.}
The fastMRI dataset~\cite{zbontar2018fastmri,knoll2020fastmri} is provided by NYU Langone. We used knee data from 227 subjects for training and 45 subjects for testing. The width and height of knee MRI data are 320 $\times$ 320.

\textbf{Hecktor.}
The Hecktor dataset~\cite{DBLP:journals/mia/OreillerAJBECVZ22} is a segmentation and outcome prediction challenge dataset about head and neck PET/CT images. We utilized the PET image to realize the PET image super-resolution. 366 subjects are used for training, and 158 subjects are used for testing. The width and height of PET data are 128 $\times$ 128.

We utilized the peak signal to noise ratio (PSNR) and structural similarity index (SSIM)~\cite{45} to evaluate the quality of super-resolved images. The PSNR is computed as the mean squared error (MSE) between the generated image $I_{SR}$ and ground truth image $I_{HR}$. The higher PSNR value indicates higher image quality. 
The SSIM was leveraged to calculate the degree of similarity between two images while accounting for elements including brightness, contrast, and structure. The value of the SSIM falls between [0, 1]. Naturally, the less image distortion there is, the higher the SSIM value.

\begin{table}[t]
\centering
\caption{Quantitative results with different methods on fastMRI and IXI dataset under 2$\times$ upsampling factor.}\label{tab1}
\setlength{\tabcolsep}{4mm}{
\begin{tabular}{c|cc|cc}
\hline
\multirow{2}{*}{Method} & \multicolumn{2}{c|}{fastMRI 2$\times$} & \multicolumn{2}{c}{IXI 2$\times$} \\  
                        & PSNR$\uparrow$           & SSIM$\uparrow$           & PSNR$\uparrow$          & SSIM$\uparrow$        \\ \hline
SRCNN~\cite{8}                   & 25.82          & 0.5602       & 29.23        & 0.8649     \\
VDSR~\cite{9}                    & 27.42          & 0.6263        & 29.79        & 0.8772     \\
FMISR~\cite{19}                   & 26.19          & 0.5583        & 29.50      & 0.8685     \\
T$^{2}$Net~\cite{12}                   & \underline{32.00}           & 0.7158         & 31.31      & 0.9035     \\
DiVANet~\cite{DBLP:journals/pr/BehjatiRFHMG23}                     & {31.98}       & \underline{0.7169}        & \underline{33.15}      & \underline{0.9320}      \\
ICONet(ours)                     & \textbf{32.12}        & \textbf{0.7191}        & \textbf{33.76}       & \textbf{0.9388}      \\ \hline
\end{tabular}}
\end{table}

\begin{table}[t]
\centering
\caption{Quantitative results with different methods on fastMRI and IXI dataset under 4$\times$ upsampling factor.}\label{tab2}
\setlength{\tabcolsep}{4mm}{
\begin{tabular}{c|cc|cc}
\hline
\multirow{2}{*}{Method} & \multicolumn{2}{c|}{fastMRI 4$\times$} & \multicolumn{2}{c}{IXI 4$\times$} \\  
                        & PSNR$\uparrow$           & SSIM$\uparrow$           & PSNR$\uparrow$          & SSIM$\uparrow$        \\ \hline
SRCNN~\cite{8}                   & 19.74          & 0.3653       & 28.12        & 0.8357     \\
VDSR~\cite{9}                    & 20.31          & 0.3839        & 28.34        & 0.8392     \\
FMISR~\cite{19}                   & 24.35          & 0.5207        & 28.27      & 0.8349     \\
T$^{2}$Net~\cite{12}                   & {30.56}           & {0.6244}         & 29.73      & 0.8773     \\
DiVANet~\cite{DBLP:journals/pr/BehjatiRFHMG23}                     & \underline{30.62 }       & \underline{0.6352}        & \underline{30.46}      & \underline{0.8946}      \\
ICONet(ours)                       & \textbf{30.69}        & \textbf{0.6385}        & \textbf{31.24}       & \textbf{0.9064}      \\ \hline
\end{tabular}}
\end{table}

\begin{table}[t]
\centering
\caption{Quantitative results with different methods on Hecktor dataset under 2$\times$ and 4$\times$ upsampling factor.}\label{tab3}
\setlength{\tabcolsep}{4mm}{
\begin{tabular}{c|cc|cc}
\hline
\multirow{2}{*}{Method} & \multicolumn{2}{c|}{Hecktor 2$\times$} & \multicolumn{2}{c}{Hecktor 4$\times$} \\  
                        & PSNR$\uparrow$           & SSIM$\uparrow$           & PSNR$\uparrow$          & SSIM$\uparrow$        \\ \hline
T$^{2}$Net~\cite{12}                   & {52.75}           & {0.9965}         & 44.16      & 0.9859     \\
ICONet(ours)                       & \textbf{53.83}        & \textbf{0.9970}        & \textbf{45.72}       & \textbf{0.9882}      \\ \hline
\end{tabular}}
\end{table}

\begin{figure*}[t]
    \centering
    \includegraphics[width=0.95\linewidth]{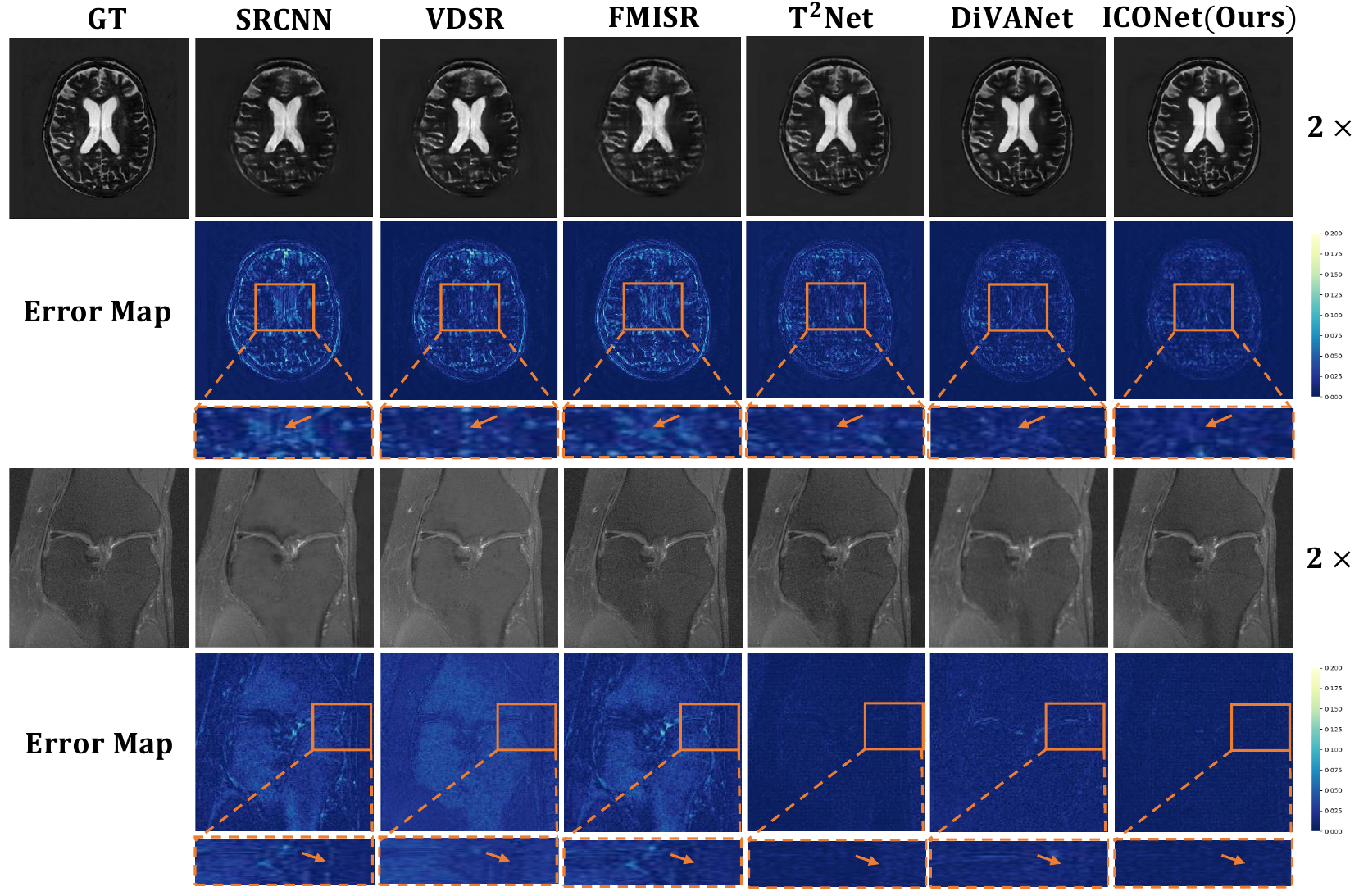}
    \caption{Qualitative results on fastMRI and IXI dataset under 2$\times$ upsampling factor. The significant differences between different methods are shown by the yellow arrow.}
    \label{fig:2-Scale-IXI-fastMRI}
\end{figure*}

\begin{figure*}[t]
    \centering
    \includegraphics[width=0.95\linewidth]{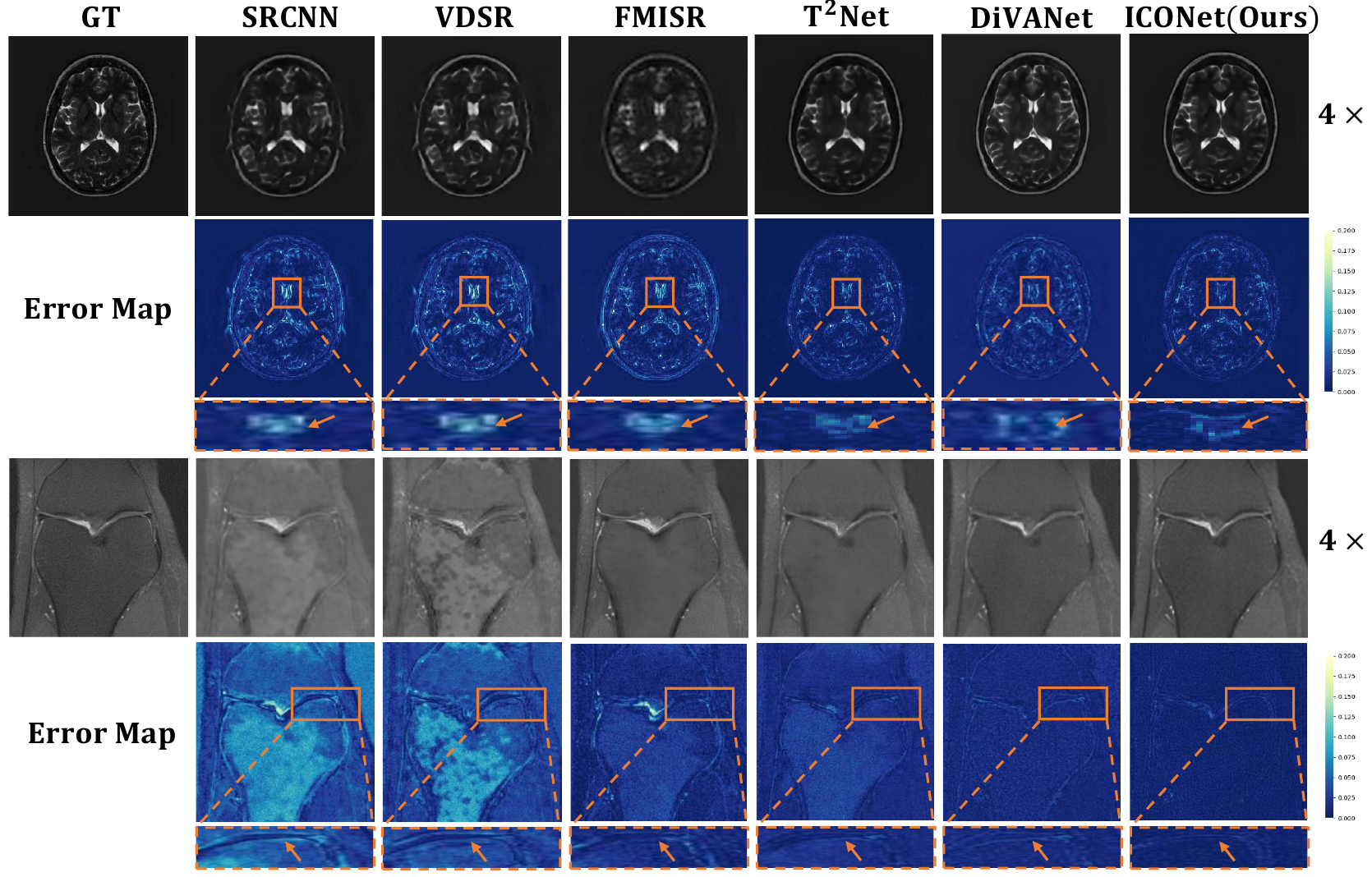}
    \caption{Qualitative results on fastMRI and IXI dataset under 4$\times$ upsampling factor. The significant differences between different methods are shown by the yellow arrow.}
    \label{fig:4-Scale-IXI-fastMRI}
\end{figure*}

\begin{figure}[t]
    \centering
    \includegraphics[width=0.85\linewidth]{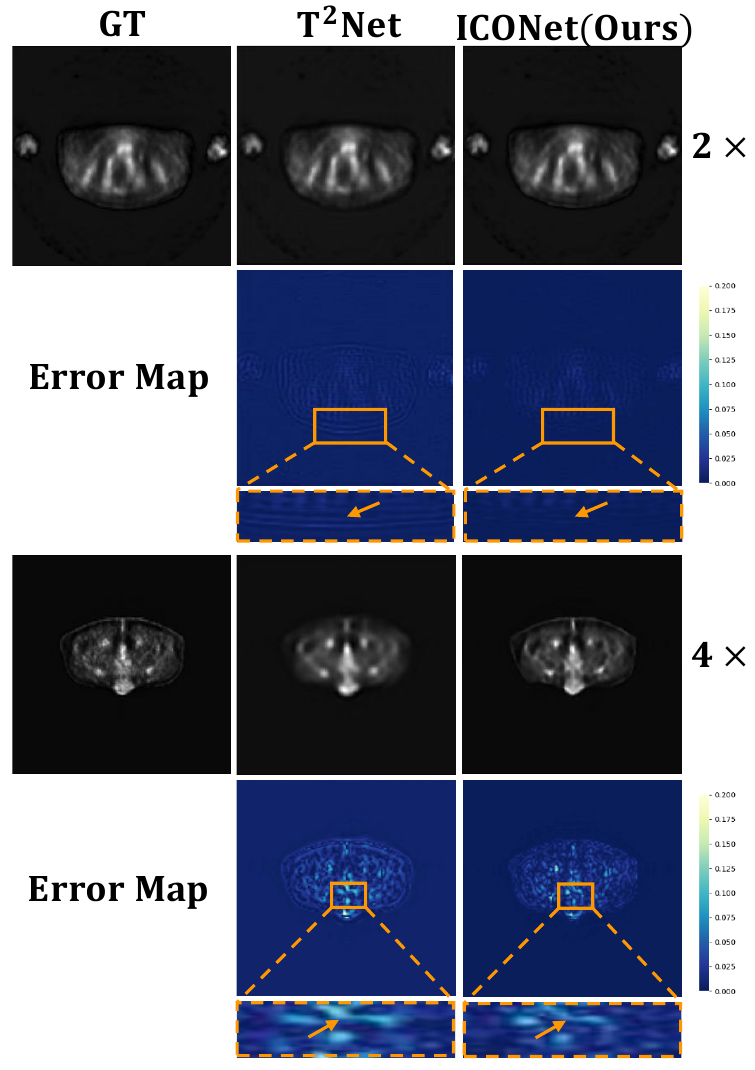}
    \caption{Qualitative results on Hecktor dataset under 2$\times$ and 4$\times$ upsampling factor. The significant differences between different methods are shown by the yellow arrow.}
    \label{fig:PETImage}
\end{figure}

\subsection{Experimental Details}
To produce LR medical images, we implement the degradation model in the frequency domain as described in \cite{36}, which helps generate LR images that more closely resemble the characteristics of real-world scenarios.
Our ICONet was implemented in PyTorch deep learning platform, and the 
experiments are performed on an NVIDIA A6000 GPU.
The Adam optimizer with a learning rate of $1 \times 10^{-4}$ was utilized to update relevant parameters. The number of feature maps was 96. The weights for $L_{SR}$ and $L_{Rec}$ are 0.9 and 0.1, respectively.

\subsection{Comparison with state-of-the-art methods}

\textbf{Quantitative Analysis.}
We compared our ICONet to several state-of-the-art methods on the IXI and fastMRI dataset under $2 \times$ and $4 \times$ upsampling factors, including SRCNN~\cite{8}, VDSR~\cite{9},  FMISR~\cite{19}, T$^2$Net~\cite{12}, and DIVANet~\cite{DBLP:journals/pr/BehjatiRFHMG23}. As the downsampling ratio increases, the loss of high-frequency information intensifies. Consequently, the challenge of MR image super-resolution becomes greater with higher upsampling factors. The quantitative comparison experiment results are shown in Tab.~\ref{tab1} and Tab.~\ref{tab2} where the best method has been highlighted. To verify the effectiveness of our ICONet on other medical modality, we also compared our approach with T$^2$Net on the Hecktor dataset (Tab.~\ref{tab3}). It can be seen from the table that the PSNR and SSIM values of our ICONet obtain the best performance than those of other methods on all upsampling factors, that is, our ICONet can effectively generate high-resolution MR images. This is because our ICONet can iteratively integrate the reconstruction prior to the super-resolution task. Furthermore, the multi-scale supervision can progressively generate super-resolved images and thus boost the SR model performance. 

\begin{table}[t]
	\centering
	\setlength{\belowcaptionskip}{0.2cm}
	\caption{Ablation study with different components of our ICONet.}
	\renewcommand\arraystretch{1.5} 
	\setlength{\tabcolsep}{1mm}{
		\label{tab4}
		\begin{tabular}{l|l|l|c|c|c}
			\hline
			Method  & SR & Rec & \makecell[c]{SR-Rec\\Fusion Module} & PSNR$\uparrow$    & SSIM$\uparrow$      \\ \hline
			\textit{Base} & \checkmark  & \ding{55}   & \ding{55}    & 33.12 & 0.9313  \\ \hline
			\textit{Base+Rec}   & \checkmark  & \checkmark   & \ding{55}    & 33.43 & 0.9348  \\ \hline
			\makecell[l]{\textit{Base+Rec+SR-Rec}\\\textit{Fusion Module (ICONet)}}  
			& \checkmark  & \checkmark   & \checkmark    & \textbf{33.76} & \textbf{0.9388}  \\ \hline
	\end{tabular}}
\end{table}

\begin{figure}[t]
    \centering
    \includegraphics[width=1\linewidth]{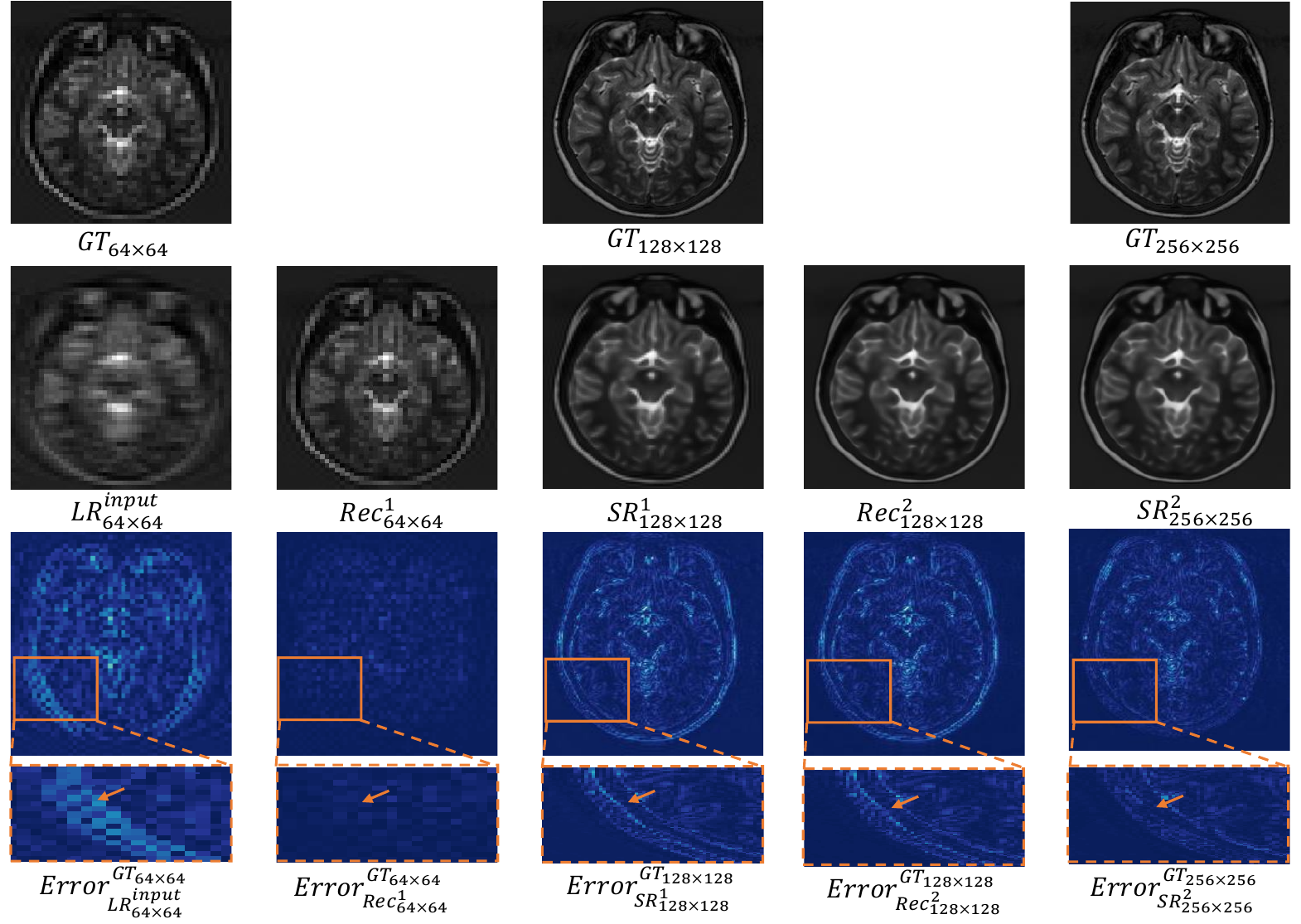}
    \caption{Artifact reduction achieved via reconstruction branch in progressive super-resolution, visualized with multi-scale error maps on the IXI dataset ($\times$4 upsampling from 64$\times$64 to 256$\times$256).}
    \label{fig:progressive_vis}
\end{figure}

\textbf{Qualitative Analysis.}
The qualitative results for $2 \times$ and $4 \times$ upsampling are shown in Fig.~\ref{fig:2-Scale-IXI-fastMRI}, Fig.~\ref{fig:4-Scale-IXI-fastMRI} and Fig.~\ref{fig:PETImage}. In addition to displaying the super-resolved image, we also present the corresponding error map. An error map is created by calculating the difference between the generated image and the high-resolution image. This allows for a clearer visualization of the discrepancies with the ground truth image. The darker the color, the smaller the difference between the super-resolved image and the ground truth image, and vice versa. The T$^2$Net~\cite{12} method, simultaneously implements super-resolution and reconstruction, and exchanges the information in a parallel way. It can obtain better results, but still, struggle to generate high-resolution images. It can be clearly seen from the red arrow in the error map that the color of our method is the darkest. Therefore, it can be proven that our ICONet can generate more accurate and detailed information in contrast to other methods.

\subsection{Ablation Study}

\textbf{1)~Effect of different modules in the iterative collaboration network:}
We studied the importance of the designed components in our ICONet architecture. Specifically, we compared the following variations: (1)~\textit{Base} module to only use the super-resolution network to super-resolve the low-resolution image, (2)~\textit{Base+Rec} module to combine the reconstruction process to the super-resolution tasks, and (3)~\textit{Base+Rec+SR-Rec fusion module(ICONet)} module effectively fuse the reconstruction information to SR by the SR-Rec fusion module. 
As shown in Table~\ref{tab4}, the \textbf{Base} model yields the lowest PSNR and SSIM values, indicating that the SR network alone is insufficient to recover fine anatomical structures and suppress artifacts effectively. Upon adding the reconstruction branch (\textbf{Base+Rec}), we observe consistent performance gains. This improvement highlights the benefit of iterative reconstruction supervision, which introduces complementary structural cues and facilitates more informed SR learning.
Finally, the full \textbf{ICONet} model incorporating the SR-Rec fusion module achieves the highest performance across all metrics. 
This highlights the value of explicit cross-task feature fusion, which enhances branch interaction and reduces information loss and error accumulation in multi-task settings.

To qualitatively demonstrate the effectiveness of our iterative collaboration network, we visualize the progressive reconstruction and super-resolution at each stage on the IXI dataset. Starting from a $64 \times 64$ low-resolution input with clearly visible artifacts, the image is first refined by the reconstruction branch to suppress artifacts and enhance structural fidelity. It is then super-resolved to $128 \times 128$, followed by another round of reconstruction and final upsampling to $256 \times 256$. The model iteratively performs reconstruction and super-resolution under the guidance of learned priors. As shown in Fig.~\ref{fig:progressive_vis}, the first row shows the ground-truth images at different resolutions, serving as supervision targets. The second row displays the model outputs at each stage: low-resolution input, initial reconstruction, intermediate super-resolution, intermediate reconstruction, and the final high-resolution result. To better show the visualization effect, the third row presents the error maps between each output and its corresponding ground-truth image. 
The results clearly show that the proposed reconstruction branch effectively reduces artifacts at early stages, while the overall framework progressively improves image quality across resolutions. 
Furthermore, the inclusion of intermediate outputs offers interpretable insights into the behavior of model and addresses the importance of intermediate supervision in multi-stage learning.
In summary, our ICONet can effectively exploit the reconstruction prior to realize medical image super-resolution.

\begin{table}[t]
	\centering
	\setlength{\belowcaptionskip}{0.2cm}
	\caption{Ablation study with serial and parallel framework under $2 \times$ upsampling factor. }
	\renewcommand\arraystretch{1.5} 
	\setlength{\tabcolsep}{10.0mm}{
		\label{tab5}
		\begin{tabular}{l|l|l}
			\hline
			Method   & PSNR$\uparrow$    & SSIM$\uparrow$     \\ \hline
			   Parallel & 33.41 & 0.9349   \\ \hline
			Serial & 33.76 & 0.9388 \\ \hline
	\end{tabular}}
\end{table}

\begin{table}[t]
	\centering
	
	\setlength{\belowcaptionskip}{0.2cm}
	\caption{Ablation study on the ratio of the super-resolution loss weight $\alpha$ and reconstruction loss weight $1 - \alpha$ in terms of PSNR and SSIM.  }
	\renewcommand\arraystretch{1.2} 
	\label{tab7}
	\setlength{\tabcolsep}{11mm}{
		\begin{tabular}{l|l|l}
			\hline
			$\alpha$ &  PSNR$\uparrow$                                    & SSIM$\uparrow$                                             \\ \hline
			
			\textbf{0.9}                      & \textbf{33.76}                                 & \textbf{0.9388}                                           \\ \hline
			0.8                        & 33.63                                 & 0.9375                                           \\ \hline
			{0.7}       &{33.51} & {0.9356}  \\ \hline
			0.6                      & 33.51                                 & 0.9359                                           \\ \hline
			0.5                        & 33.47                                 & 0.9355                                           \\ \hline
	\end{tabular}}
\end{table}

\begin{table}[t]
\centering
\caption{Ablation study on using different numbers of residual spatial-channel feature learning (RSCFL) modules.}
\begin{tabular}{p{2cm}|p{2cm}|p{2cm}}
\hline
\centering\textbf{Number} & \centering\textbf{PSNR} & \centering\textbf{SSIM} \tabularnewline
\hline
\centering 3 & \centering 32.91 & \centering 0.9290 \tabularnewline
\centering 4 & \centering 33.19 & \centering 0.9323 \tabularnewline
\centering 5 & \centering 33.41 & \centering 0.9349 \tabularnewline
\centering 6 & \centering 33.51 & \centering 0.9356 \tabularnewline
\centering 7 & \centering 33.74 & \centering 0.9386 \tabularnewline
\hline
\end{tabular}
\label{tab:rscfl_ablation}
\end{table}

\begin{table}[t]
	\centering
	\setlength{\belowcaptionskip}{0.2cm}
	\caption{Ablation study with single-stage and multi-stage supervision comparisons under $4 \times$ upsampling factor. }
	\renewcommand\arraystretch{1.5} 
	\setlength{\tabcolsep}{9.0mm}{
		\label{tab8}
		\begin{tabular}{l|l|l}
			\hline
			Method   & PSNR$\uparrow$    & SSIM$\uparrow$    \\ \hline
			ICONet-SS   & 31.02 & 0.9022  \\ \hline
			ICONet-MS & 31.24 & 0.9064 \\ \hline
	\end{tabular}}
\end{table}

\begin{table}[t]
	\centering
	\setlength{\belowcaptionskip}{0.2cm}
	\caption{Ablation study with vision Mamba and SwinIR on model complexity. }
	\renewcommand\arraystretch{1.5} 
	\setlength{\tabcolsep}{2.0mm}{
		\label{tab6}
		\begin{tabular}{l|l|l|l|l}
			\hline
			Method   & Params[M]    & FLOPS[G]  & PSNR$\uparrow$ & SSIM$\uparrow$ \\ \hline
			vision Mamba   & 0.996 & 9.61 &  33.05&0.9304 \\ \hline
			SwinIR~\cite{DBLP:conf/iccvw/LiangCSZGT21} & 1.260 & 20.66 & 32.62&0.9253 \\ \hline
	\end{tabular}}
\end{table}

\textbf{2)~Discussions about the serial and parallel frameworks:} 
To evaluate the effectiveness of the serial and parallel framework, we test two models in the ablation study. The first is a serial framework, which first realizes medical image reconstruction, and then the reconstruction result is used as the input of the SR model to further guide the generation of high-resolution images. The second is a parallel framework, which develops a multi-task learning network for joint super-resolution and reconstruction based on the shared features. Tab.~\ref{tab5} tabulates the super-resolution results. 
We see that the serial framework obtains a higher performance than parallel frameworks. Sequentially decoupling the reconstruction and SR stages while preserving task dependency proves effective. This design enables more targeted use of structural priors and enhances super-resolution fidelity.
While the black box-like sharable features of parallel frameworks cannot explicitly exploit the correlation of these two tasks, leading to degraded performance.

\textbf{3)~Effect of different weight combinations of loss function:}
We investigated different weight combinations between the super-resolution and reconstruction tasks, ensuring their sum equals 1. Among the tested values of $\alpha = \left\{ 0.9,0.8,0.7,0.6,0.5 \right\}$, the setting $\alpha = 0.9$ consistently achieved the best results across all evaluation metrics, as shown in Tab.~\ref{tab7}.

\textbf{4)~Discussions the number of residual spatial-channel feature learning (RSCFL) module:}
The number of residual spatial-channel feature learning (RSCFL) modules can be flexibly adjusted in our ICONet. To investigate the impact of the number $N$ of RSCFL, we compared the experimental results when we change the values of $N$. As shown in Table~\ref{tab:rscfl_ablation}, increasing $N$ generally improves PSNR and SSIM, indicating that deeper spatial-channel modeling helps capture more complex structural patterns. We adopt $N = 7$ in our final model to achieve strong performance while maintaining a good trade-off with computational efficiency.

\textbf{5)~Effect of multi-stage supervision:}
To illustrate the effectiveness of the multi-stage supervision, we created two baselines as follows: 1) ICONet-SS: The ICONet model, trained with single-stage supervision. Specifically, the supervision only on the output scale; 2) ICONet-MS: The ICONet model with multi-stage supervision, which means the supervision on each stage of the network. The experimental results are shown in Tab.~\ref{tab8}. 
As shown in Table~\ref{tab8}, ICONet-MS consistently outperforms ICONet-SS, achieving a PSNR improvement from 31.02 dB to 31.24 dB and a SSIM gain from 0.9022 to 0.9064. 
This indicates that multi-stage supervision provides more informative gradient signals throughout the network, stabilizing intermediate predictions and reducing error propagation.

\textbf{6)~Baseline selection:}
We also conducted the experiment about the baseline selection to validate the effectiveness of vision Mamba and Transformer-based method. Specifically, vision Transformer has high computational complexity when processing high-resolution images because the computational complexity of the self-attention mechanism grows quadratically with the number of image patches. Swin Transformer~\cite{DBLP:conf/iccv/LiuL00W0LG21} effectively reduces computational complexity through its window and hierarchical mechanisms. Liang~\cite{DBLP:conf/iccvw/LiangCSZGT21} first applied the Swin Transformer to the image super-resolution task. We show the parameters and FLOPs with the 128 $\times $128 input on the IXI dataset. It can be seen from Tab. \ref{tab6} that the vision Mamba achieves high-quality super-resolution with extremely few parameters and FLOPs compared with swinIR. This demonstrates that exploring vision Mamba is the right approach for achieving efficient super-resolution.

Beyond its computational efficiency, Mamba is better aligned with the structural modeling demands of medical images, particularly in restoring anatomical details and maintaining spatial continuity within tissues. Medical images (e.g., MRI) often exhibit strong directionality and local regularity, such as at gray-white matter interfaces or tumor boundaries—features essential for clarity and downstream performance.
While traditional Transformers possess powerful global modeling capabilities, their attention mechanisms are fundamentally based on content similarity between patches rather than explicit spatial relationships. They lack inherent awareness of structural continuity and directional context. Although some variants introduce positional encodings, their ability to model spatial structure remains limited in high-resolution scenarios. 
This may cause blurred boundaries and artifacts by mislinking visually similar but spatially distant regions, which is especially harmful in complex anatomical areas.
In contrast, the Mamba module incorporates selective scanning and recurrent update mechanisms to enable a sequential modeling process across the spatial domain. This design naturally preserves spatial order and enhances the ability to capture structural continuity and edge directionality. It aligns well with the characteristics of medical images, where structural extension and localized dependencies are prominent. Therefore, 
Mamba more effectively delineates anatomical boundaries and preserves tissue consistency, while also improving sensitivity to pathological regions and ensuring high-fidelity super-resolution.

\begin{figure}[t]
    \centering
    \includegraphics[width=1\linewidth]{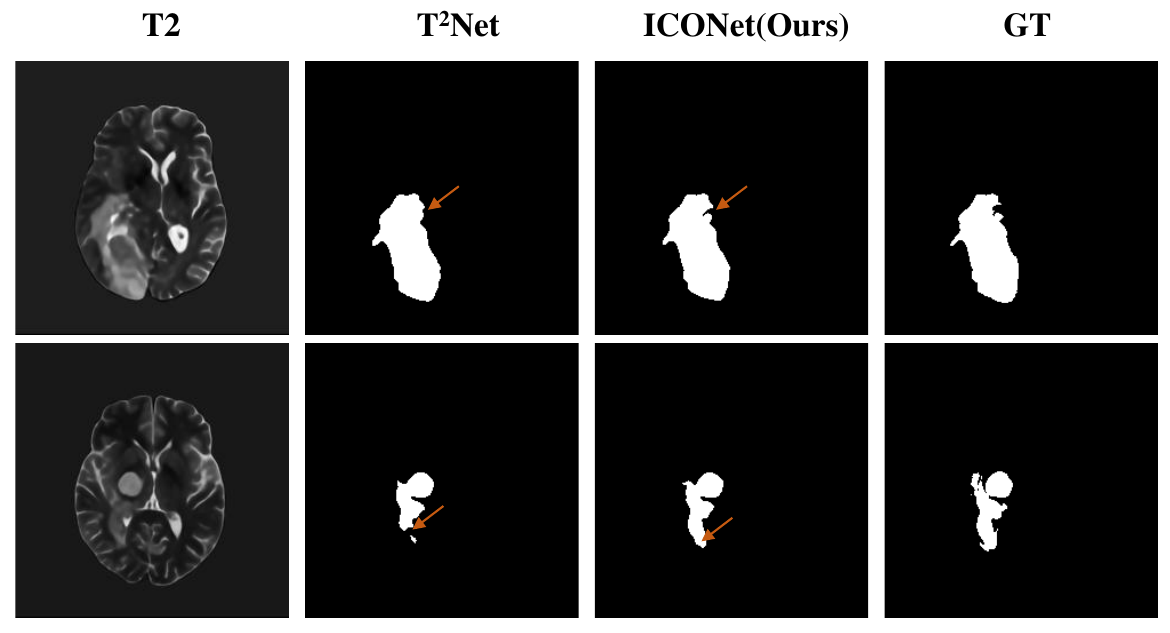}
    \caption{Visual comparison of segmentation results using SR inputs generated by T$^2$Net and our ICONet. ICONet provides clearer tumor boundaries and structural details.}
    \label{fig:Seg}
\end{figure}

\begin{table}[t]
    \centering
    \setlength{\belowcaptionskip}{0.2cm}
    \caption{Segmentation performance comparison (Dice Score) using super-resolved images from different methods.}
    \renewcommand\arraystretch{1.5} 
    \setlength{\tabcolsep}{7.0mm}{
    \label{tabseg}
    \begin{tabular}{l|l}
        \hline
        Method & Dice Score$\uparrow$ \\
        \hline
        T$^{2}$Net & 0.5019 \\
        ICONet(ours) & 0.5265 \\
        \hline
    \end{tabular}}
\end{table}

\subsection{Application to Brain Tumor Segmentation Task}

To evaluate the clinical task performance of our ICONet, we conducted a downstream brain tumor segmentation experiment on the BraTS2021\footnote{http://braintumorsegmentation.org/} dataset. Accurate tumor segmentation is crucial for radiotherapy, as it directly determines how the radiation dose is planned and delivered to the tumor. We focused on the whole tumor segmentation using 2D slices.
Specifically, we performed super-resolution on T2 images using both T$^2$Net and our ICONet. To simulate a practical diagnostic setting, the high-resolution T1 images were additionally provided as auxiliary inputs for segmentation. The segmentation task was implemented using the Attention U-Net~\cite{DBLP:journals/corr/abs-1804-03999} as the backbone model.
As shown in Tab.~\ref{tabseg}, ICONet achieved a higher Dice Score of 0.5265 compared to 0.5019 from T$^2$Net. In addition, Fig.~\ref{fig:Seg} qualitatively illustrates that ICONet yields more accurate tumor boundaries and better structural consistency. This performance gain demonstrates that our method not only generates high-quality images but also enhances the effectiveness of downstream tumor segmentation, highlighting its practical diagnostic value.

\section{Conclusion and Future Work}\label{Conclusion}

In this paper, we have developed an iterative collaboration network~(ICONet) for medical image super-resolution. Specifically, the super-resolution module makes use of the prior knowledge of reconstruction to iteratively produce high-resolution images, enabling more precise reconstruction in turn. 
The designed residual spatial-channel feature learning (RSCFL) module can efficiently mine effective features beneficial for super-resolution in spatial and channel dimensions.
A SR-Rec fusion module is designed to improve performance by strengthening the guidance of the reconstruction prior, where artifact-free features are attentively collected. Furthermore, we also develop the multi-stage loss to more effectively constrain the SR training process and achieve higher upsampling. Experimental results, both quantitative and qualitative, demonstrate that our ICONet greatly outperforms state-of-the-art medical image super-resolution approaches.

The proposed ICONet has a few potential limitations as well, which will be addressed in our follow-up work. Our ICONet only considers medical image super-resolution in the image domain, while frequency data can also provide original and effective information. ICONet could be extended to both the image domain and frequency domain, which is also an important direction for future work.

\bibliographystyle{IEEEtran}
\bibliography{mybibfile.bib}

\end{document}